\documentclass[a4paper,11pt]{article}
\usepackage{pos}
\usepackage{amsmath}
\usepackage{tabularx}
\usepackage{multirow}
\usepackage{float}
\newcommand{\qvec}{\boldsymbol{q}}

\title{Electric Polarizability of Charged Pions from nHYP Four-Point Functions}

\author*[a]{Benjamin Luke}
\author[a]{Sudip Shiwakoti}
\author[a]{Shayan Nadeem}
\author[b]{Andrei Alexandru}
\author[a]{Walter Wilcox}
\author[b]{Frank Lee}

\affiliation[a]{Department of Physics and Astronomy, Baylor University,\\
  Waco, Texas 76798,USA}

\affiliation[b]{Physics Department, The George Washington University,\\
Washington, District of Columbia 20052,USA}

\emailAdd{Ben\_Luke1@baylor.edu}
\emailAdd{Sudip\_Shiwakoti1@baylor.edu}
\emailAdd{Shayan\_Nadeem1@baylor.edu}
\emailAdd{aalexan@gwu.edu}
\emailAdd{Walter\_Wilcox@baylor.edu}
\emailAdd{fxlee@gwu.edu}

\abstract{Understanding a hadron’s electric and magnetic polarizabilities allows one to access internal structural information. Traditionally, the external field two-point function method has been used to calculate polarizabilities. However, recent work has demonstrated the effectiveness of using four-point functions for computing polarizabilities of charged and neutral hadrons. Our previous study on the electric polarizability of the charged pion used a quenched Wilson action on a lattice with pion mass from 1100 MeV to 370 MeV. In this work, we employ a number of improvements, including a dynamical action (nHYP), smaller pion masses (220 MeV and 315 MeV), and a variable lattice size in order to extrapolate to infinite volume. Preliminary results are presented.}

\FullConference{The 42nd International Symposium on Lattice Field Theory (LATTICE2025)\\
2-8 November 2025\\
Tata Institute of Fundamental Research, Mumbai, India\\}

\begin{document}
\maketitle

\section{Introduction}
The study of electromagnetic polarizabilities has been a long-standing project in hadronic physics. Computing polarizabilities on the lattice presents opportunities for novel results but also comes with serious challenges, especially when dealing with charged hadrons. Historically, the background field method has been used to compute polarizabilities of neutral hadrons in lattice QCD. However, charged hadrons present unique difficulties because they accelerate in an electric field and exhibit Landau levels in a magnetic field. These motions are unrelated to the polarizability and make the extraction of the polarizability from other information challenging.

The approach presented here circumvents the problems faced by the background field method by using four-point functions to calculate the electric polarizability of charged pions. This work is an extension of a previous study on charged pions \cite{Lee014512}. In Ref. \cite{Lee014512}, Lee et al. demonstrate the use of four-point functions for calculating electric polarizabilites of charged pions. They use a quenched Wilson action on a $24^3 \times 48$ lattice with four pion masses ranging from 1100 MeV to 370 MeV. The work presented here improves upon \cite{Lee014512} for a few reasons, all owing to the use of nHYP configurations \cite{Niyazi094506}. First, nHYP uses dynamical fermions whereas the Wilson action is quenched. Second, nHYP has two smaller pion masses (315 and 220 MeV), whereas \cite{Lee014512} uses four larger masses. Third, nHYP has a volume dependence which allows for extrapolation to infinite volume.

The goal of this paper is to report on the analysis that was done to obtain polarizability results using the nHYP configurations and compare them with those obtained from \cite{Lee014512}. It is important to mention that there is not a great deal of expectation that the results from this study agree exactly with \cite{Lee014512} since we use a different action. Nonetheless, approximate agreement is expected (and observed). The narrative around these results will continue to evolve as the analysis progresses and the results are refined.

\section{The charged pion}

\begin{figure}[h] 
   \centering
   \includegraphics[width=2.5in]{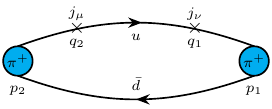} 
   \caption{Topological diagram of the four-point function for the charged pion. Time moves from right to left. The four-momentum conservation is $p_2 = p_1 + q_1 + q_2$.}
   \label{fig:4ptPionDiag}
\end{figure}

In Ref. \cite{Wilcox034506}, an equation for the electric polarizability of the charged pion on the lattice is derived (also presented in Ref. \cite{Lee014512} as Eq.(1)). It takes the form
\begin{equation} \label{eq:pol_tot}
\alpha_E^\pi = \alpha \left\{\frac{\langle r^2\rangle}{3 m_\pi} + \lim_{\qvec\rightarrow 0}\frac{2}{\qvec^2}\int_0^\infty dt \left[Q_{44}(\qvec,t)-Q_{44}^\text{elas}(\qvec,t)\right]\right\},
\end{equation}
where $\alpha = 1/137$ is the fine structure constant, $Q_{44}$ is the total four-point function, and $Q_{44}^\text{elas}$ is the elastic contribution to the four point function. The first term we will denote as the elastic contribution to the polarizability,
\begin{equation} \label{eq:pol_elas}
\alpha_E^\text{elas} = \frac{\alpha\langle r^2\rangle}{3m_\pi},
\end{equation}
and the second term we denote as the inelastic contribution,
\begin{equation} \label{eq:pol_inelas}
\alpha_E^\text{inelas} = \lim_{\qvec\rightarrow 0}\frac{2\alpha}{\qvec^2}\int_0^\infty dt \left[Q_{44}(\qvec,t)-Q_{44}^\text{elas}(\qvec,t)\right].
\end{equation}
We note that although the integral extends to $\infty$, in the lattice calculation one can only integrate across the available time slices between the two insertion points. 

We work with a special kinematic setup, called a zero-momentum Breit frame, which mimics low-energy Compton scattering (see Fig. \ref{fig:4ptPionDiag}). The initial and final pions are at rest and the photons have purely spacelike momentum,
\begin{equation}
p_1 = (\boldsymbol{0},m_\pi), \quad p_2 = (\boldsymbol{0},m_\pi), \quad q_1 = (\qvec,0), \quad q_2 = (-\qvec, 0).
\end{equation}

\begin{figure}[h] 
   \centering
   \includegraphics{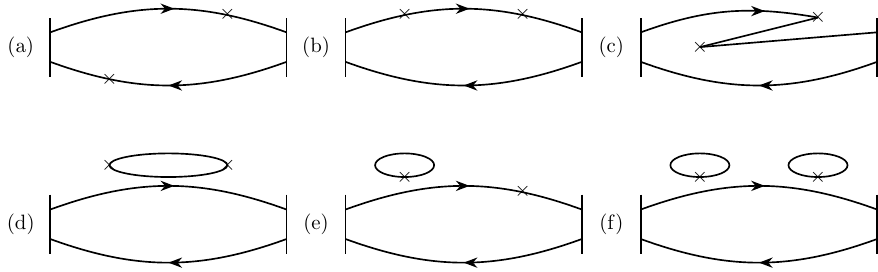} 
   \caption{All possible diagrams corresponding to the $\pi^+$ four-point function.}
   \label{fig:diagsAtoF}
\end{figure}

The complete calculation is accomplished by evaluating all possible topological diagrams, of which there are six. (See Fig. \ref{fig:diagsAtoF}.) However, given the complicated nature of evaluating the disconnected contributions, we choose to evaluate only the connected contributions (diagrams (a), (b), and (c)), and leave the disconnected calculations for future work. Note that Lee et al. also make this choice (Ref. \cite{Lee014512}).

\section{Calculation details}
We work with four nonzero momenta, $(0,0,1)$, $(0,1,1)$, $(1,1,1)$, and $(0,0,2)$, for each ensemble. The ensemble descriptions are outlined in Table \ref{table:ensDesc}. The total polarizability result is the sum of elastic and inelastic contributions, $\alpha_E^\pi = \alpha_E^\text{elas} + \alpha_E^\text{inelas}$. The basic procedure of the calculation is to compute the two summands, given by Eqs. (\ref{eq:pol_elas}) and (\ref{eq:pol_inelas}). We proceed by first evaluating the elastic contribution. 

\begin{table}[h] \centering
\begin{tabular}{cccccccc}
\hline\hline
Ensemble & $N_\text{configs}$ & Mass (MeV) & $L_x \times L_y \times L_z \times L_t$ & $a$ (fm) &  $T_0$ & $T_1$ & $T_3$ \\ \hline
1        & 300 & 315 & $24^3\times48$                       & 0.1210(2)(24)  & 6     & 17    & 41    \\
2        & 300 & 315 & $30 \times 24^2 \times 48$      &       & 6     & 17    & 41    \\
3        & 300 & 315 & $48 \times 24^2 \times 48$       &      & 6     & 17    & 41    \\
4        & 400 & 220 & $24^3 \times 64$                & 0.1215(3)(24)       & 8     & 19    & 55    \\
5        & 400 & 220 & $28\times 24^2 \times 64$        &      & 8     & 19    & 55    \\
6        & 400 & 220 & $32 \times 24^2 \times 64$        &     & 8     & 19    & 55    \\ \hline\hline
\end{tabular}
\caption{Ensemble descriptions.}
\label{table:ensDesc}
\end{table}

\subsection{Elastic contribution}
The elastic contribution to the polarizability $\alpha_E^\text{elas}$ is proportional to the charge radius squared, $\langle r^2\rangle$ (Eq.(\ref{eq:pol_elas})). The charge radius squared is determined by the pion form factor via
\begin{equation} \label{eq:r2}
\langle r^2 \rangle = -6 \frac{dF_\pi(\qvec^2)}{d\qvec^2}\bigg|_{\qvec^2\rightarrow 0}.
\end{equation}
The form factor $F_\pi$ can be obtained from its relationship with the elastic contribution to the four-point function \cite{Lee014512},
\begin{equation} \label{eq:q44_elas}
Q_{44}^\text{elas}(\qvec,t) = \frac{(E_\pi + m_\pi)^2}{4E_\pi m_\pi}F_\pi^2(\qvec^2)e^{-a(E_\pi - m_\pi)t}.
\end{equation}
The calculation starts by fitting the normalized four-point function data at large time separations to an exponential form according to Eq.(\ref{eq:q44_elas}).

\begin{figure}[h] 
   \centering
   \includegraphics[width=1.9in]{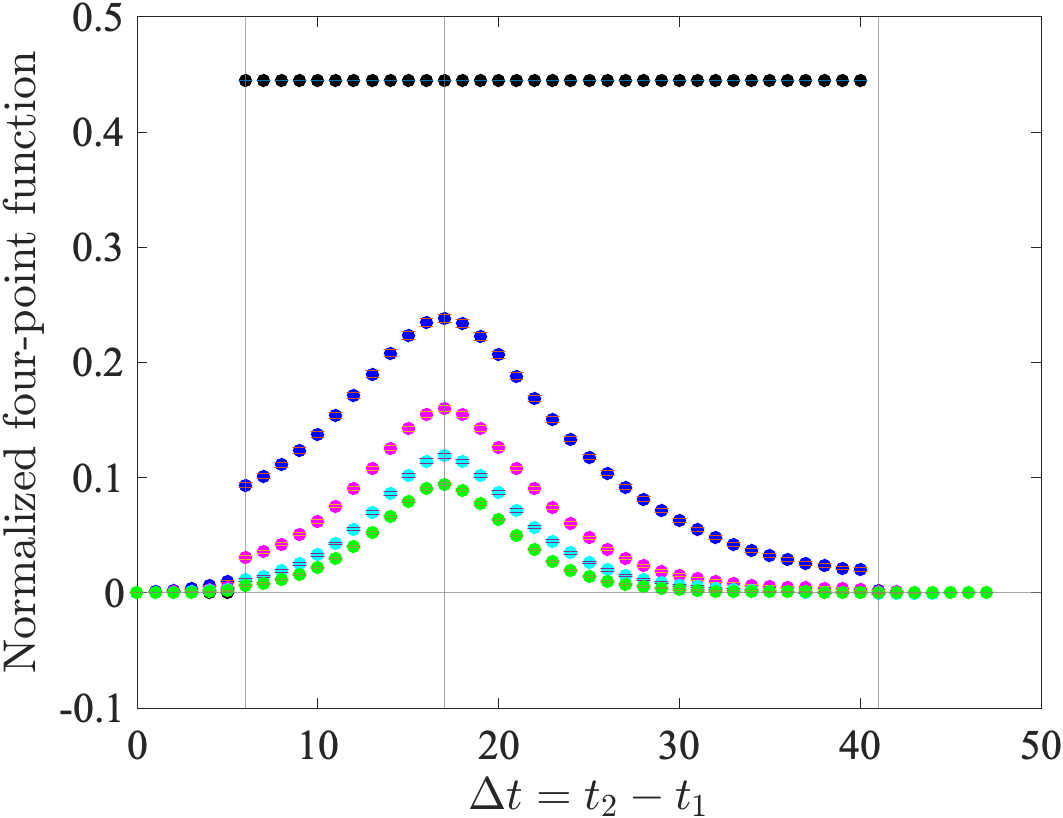} 
   \includegraphics[width=1.9in]{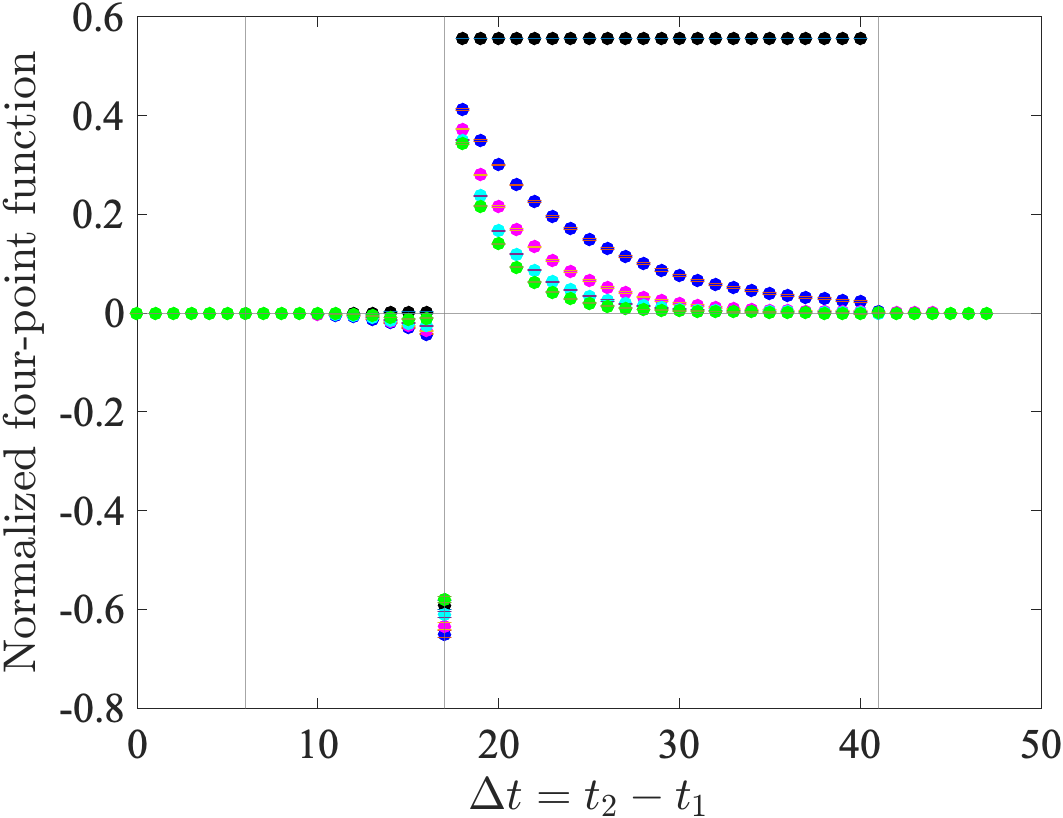} 
   \includegraphics[width=1.9in]{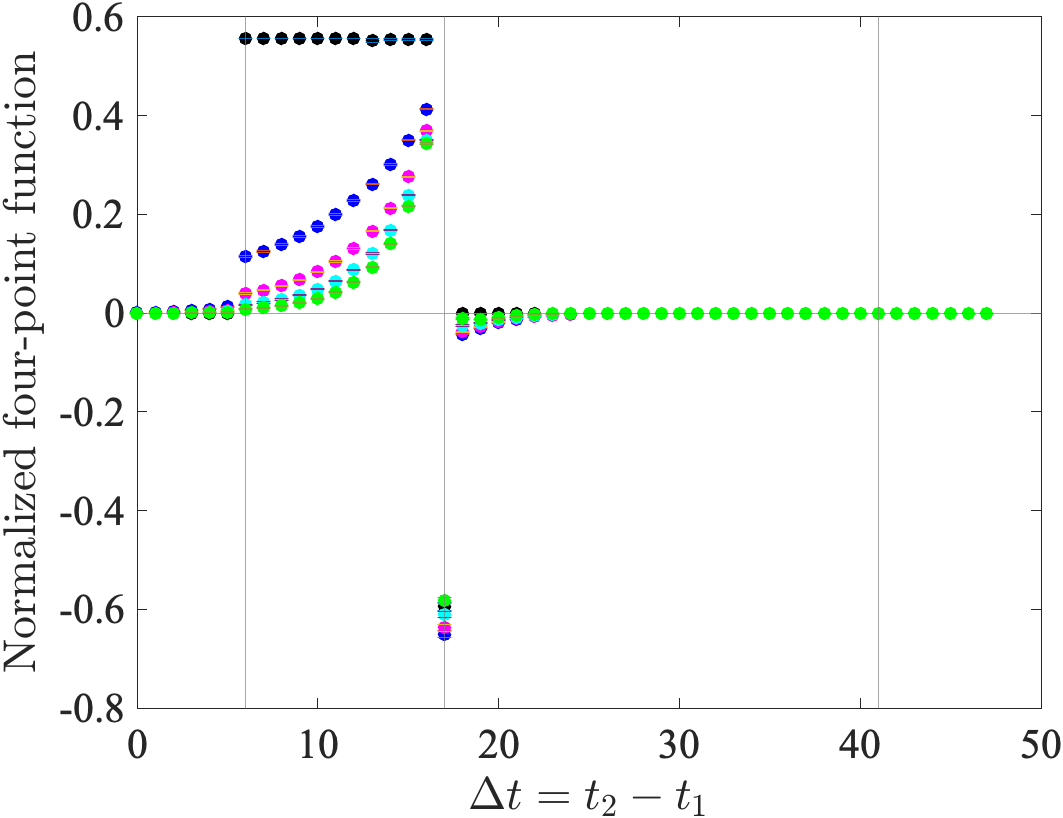} 
   \caption{Normalized four-point functions for diagrams (left to right) (a), (b), and (c) as a function of current separation. Black, blue, magenta, cyan, and green data points represent momenta (0,0,0), (0,0,1), (0,1,1), (1,1,1), and (0,0,2), respectively. The quark zero-momentum wall sources are at $t_0 = 6$ and $t_3 = 41$, with a fixed current insertion at $t_1 = 17$, as indicated by the vertical lines. The diagrams correspond only to Ensemble 1.}
   \label{fig:4ptRaw}
\end{figure}

Fig. \ref{fig:4ptRaw} gives a sample of the normalized four-point functions for the connected diagrams. One can observe that diagrams (a) and (b) exhibit an exponential behavior at large time separations $t_2 - t_1$. Diagrams (a) and (b) are responsible for the bulk of the elastic contribution and the sum of these diagrams at large time separations is what is fit according to Eq.(\ref{eq:q44_elas}).

\begin{figure}[h] 
   \centering
   \includegraphics[width=1.9in]{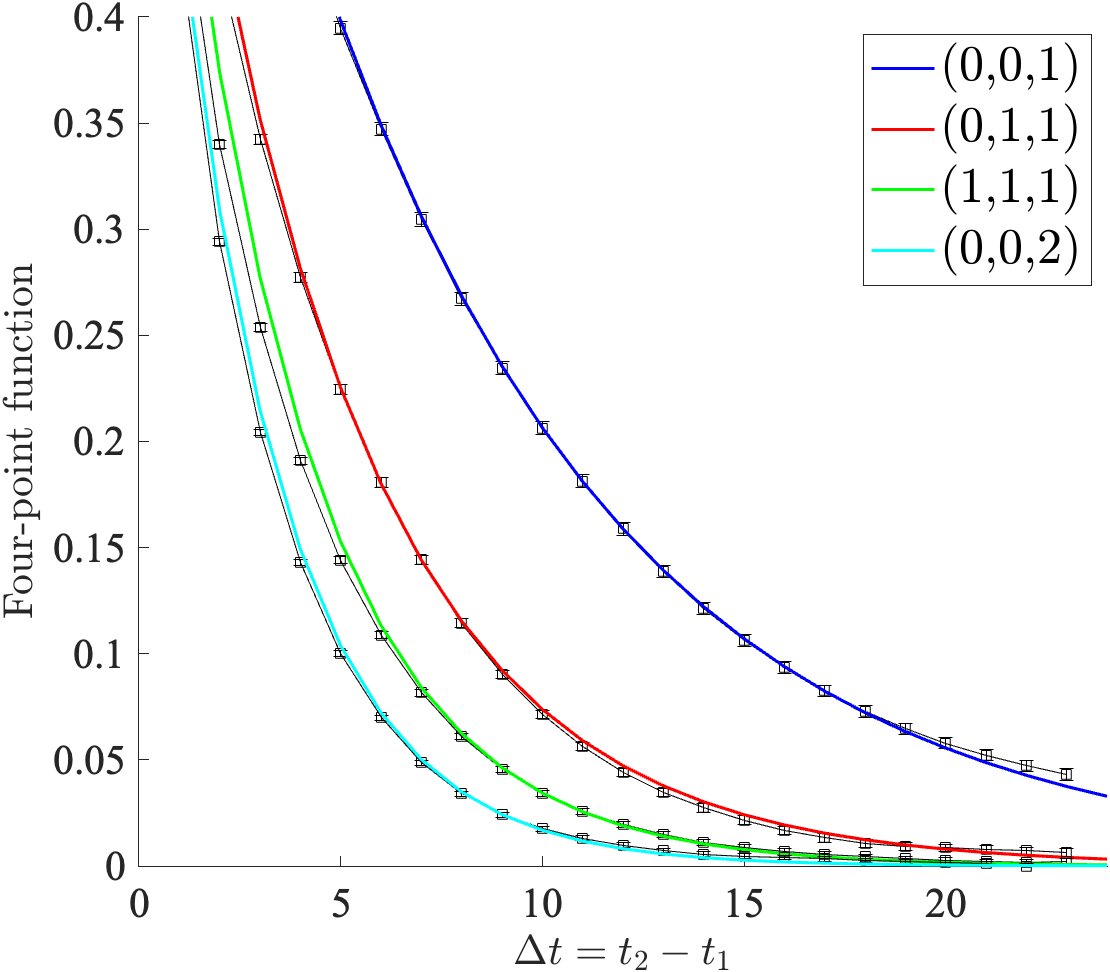} 
   \includegraphics[width=1.9in]{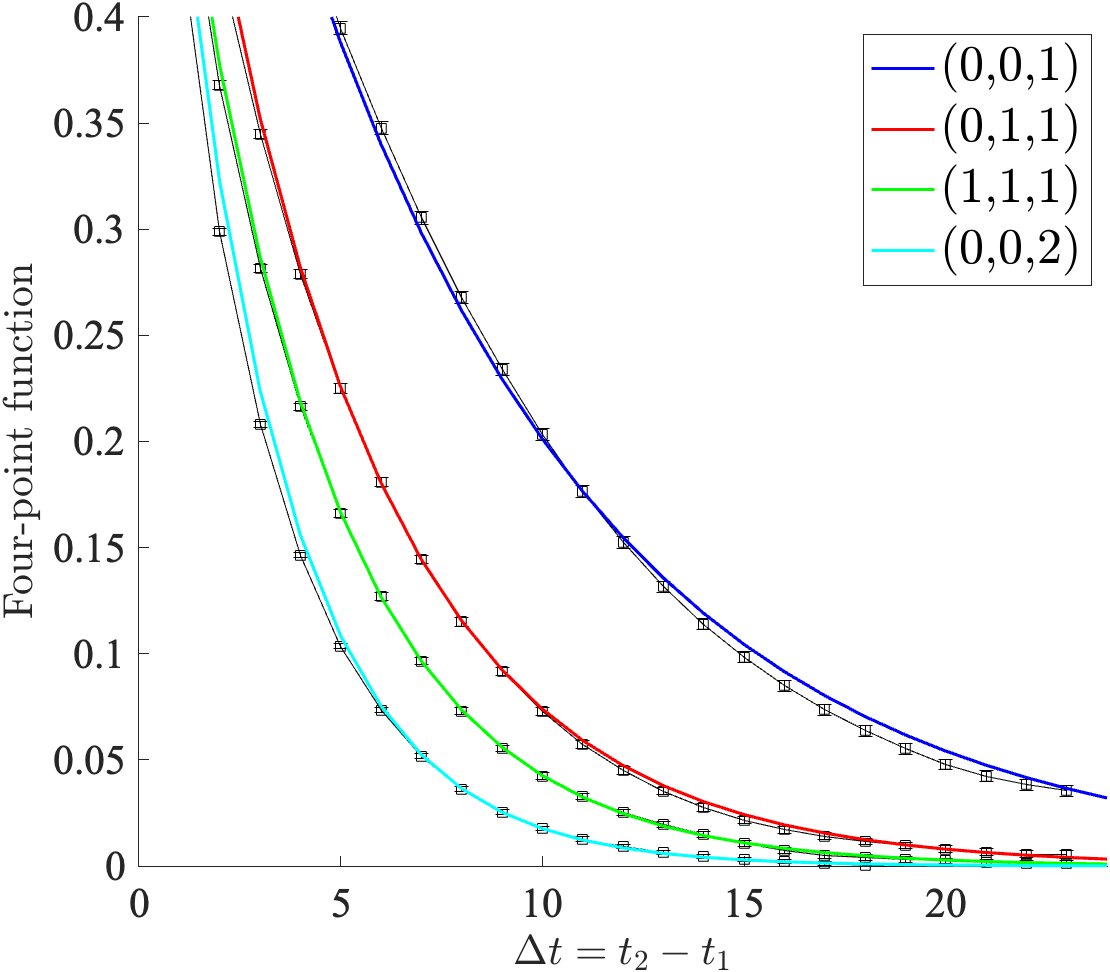} 
   \includegraphics[width=1.9in]{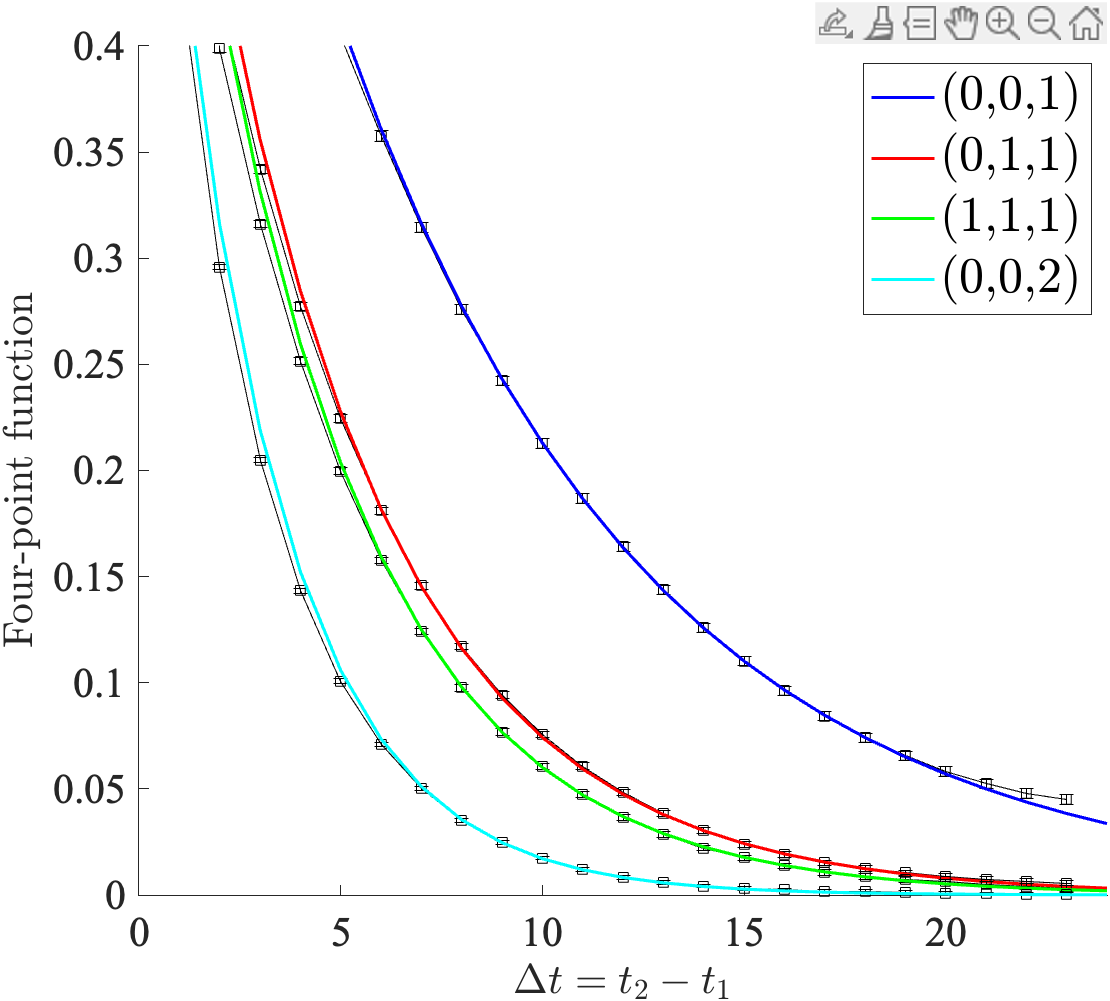} 
   \caption{Elastic fits for Ensembles 1, 2, and 3 ($m_\pi = 315$ MeV). The colored curves are the computer-generated fits and the black points represent the raw data. Note that the fits are made on the sum of diagrams (a) and (b) only.}
   \label{fig:elasFits}
\end{figure}

Fig. \ref{fig:elasFits} gives a sample of the fitted four-point functions for Ensembles 1, 2, and 3. From these fits one can solve for the form factor as a function of $\qvec^2$. This yields four data points per ensemble corresponding to the four nonzero momenta. We take these points and then fit them to obtain a form for $F_\pi(\qvec^2)$. 

There are a couple of functional forms one may choose from to fit the form factor data, including the monopole form $F_\pi(\qvec^2) = (1 + \qvec^2/m_v)^{-1}$ and the z-expansion \cite{Lee013013}. The monopole form was found to fit the data poorly, so we opted for the z-expansion with two parameters,\footnote{The z-expansion with three fit parameters was also tested, but again fit the data poorly.}
\begin{equation} \label{eq:z_exp}
\begin{gathered}
F_\pi(\qvec^2) = 1 + a_1 z + a_2 z^2, \\
\text{where} \quad z \equiv \frac{\sqrt{t_\text{cut} - t} - \sqrt{t_\text{cut} - t_0}}{\sqrt{t_\text{cut} - t} + \sqrt{t_\text{cut} - t_0}}, \\
t = -\qvec^2, \\
t_\text{cut} = 4m_\pi^2
\end{gathered}
\end{equation}

\begin{figure}[h] 
   \centering
   \includegraphics[width=1.9in]{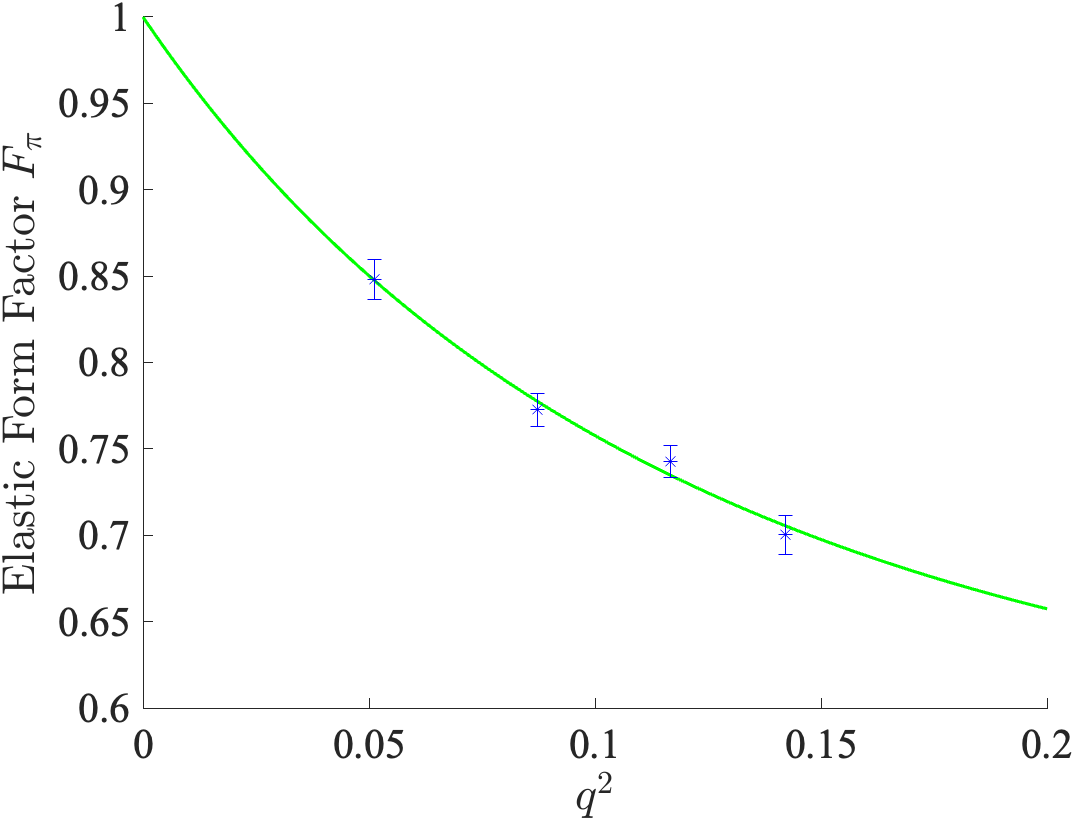} 
   \includegraphics[width=1.9in]{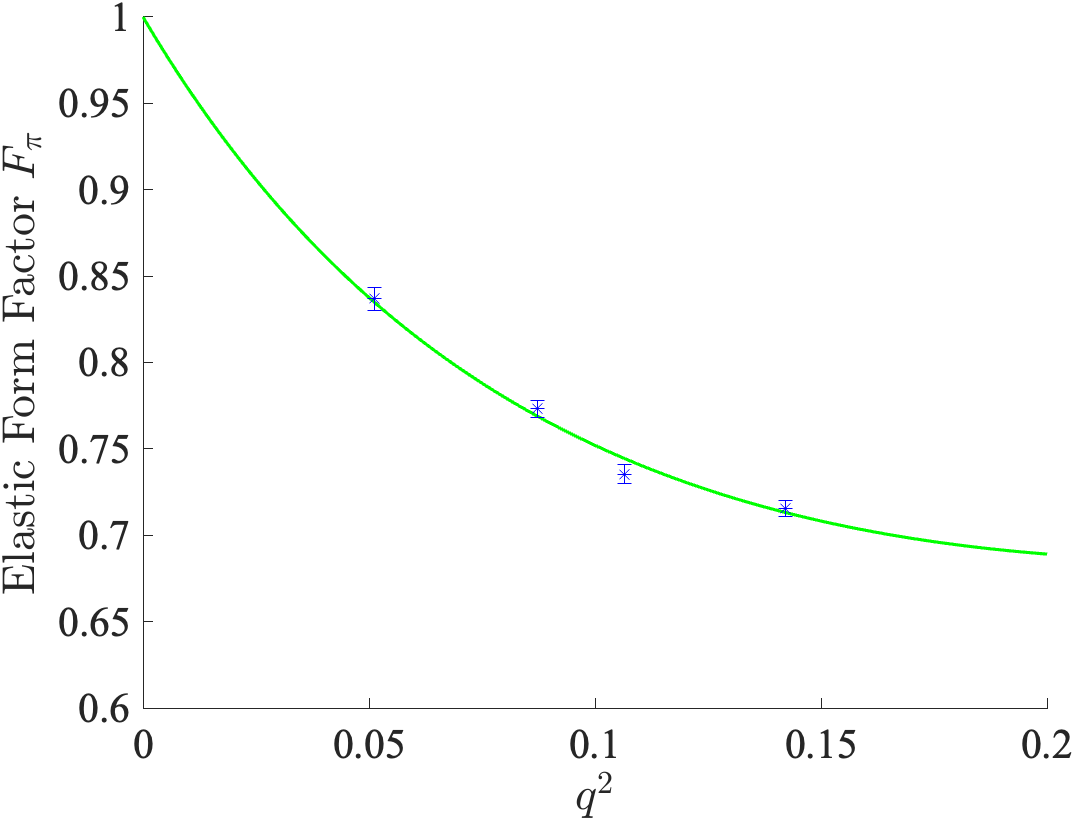} 
   \includegraphics[width=1.9in]{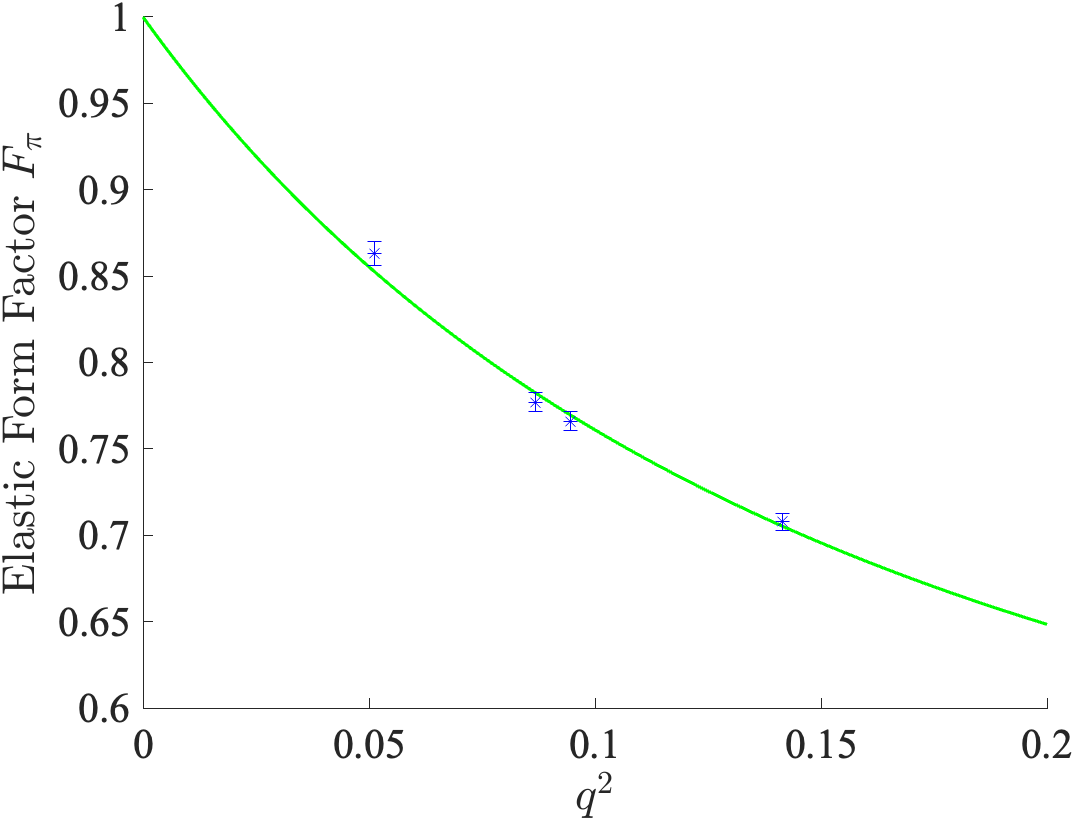} 
   \caption{Form factors for Ensembles 1, 2, and 3 ($m_\pi = 315$ MeV). A z-expansion fit with 2 parameters was used.}
   \label{fig:FF}
\end{figure}

We take $t_0 = 0$ so that $F_\pi(0) = 1$ by construction. A sample of the z-expansion fits over the form factors are shown in Fig. \ref{fig:FF}. The central difference formula\footnote{One can also compute the analytical derivative given the known form of $F_\pi$. This result was found to be identical.} is used to find the derivative at $\qvec^2 \rightarrow 0$, thus giving $\langle r^2 \rangle$ from Eq.(\ref{eq:r2}).

\begin{figure}[h] 
   \centering
   \includegraphics[width=2.8in]{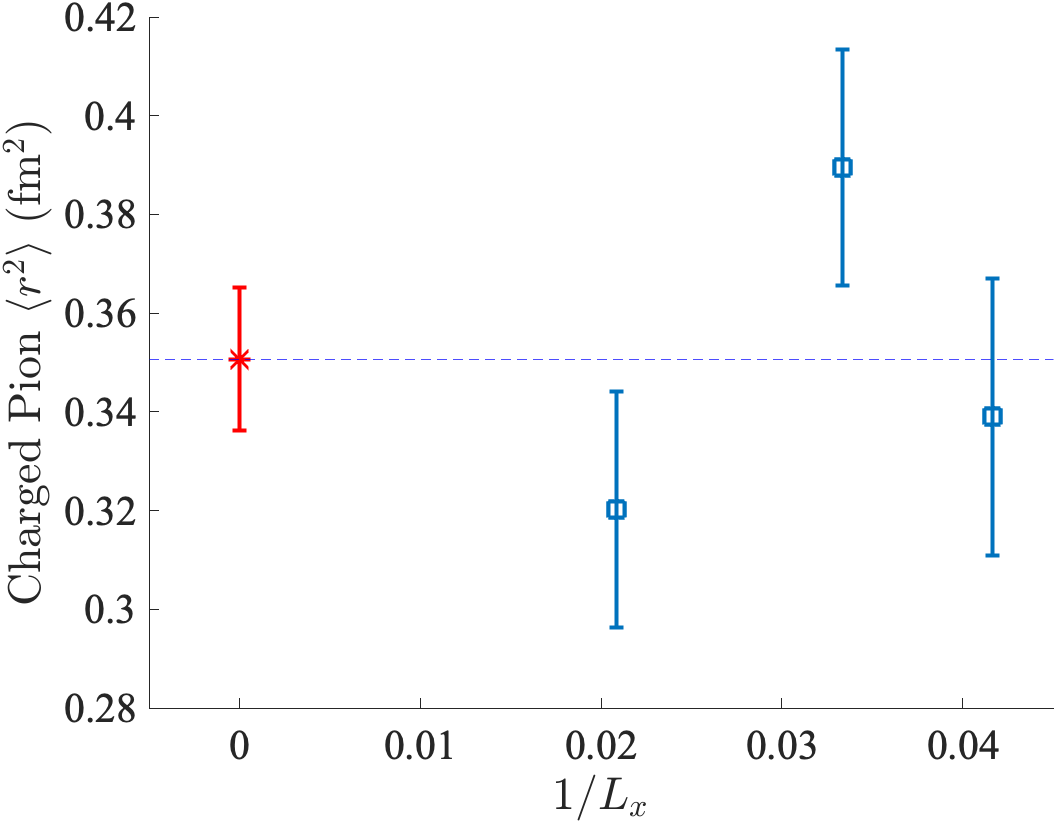} 
   \includegraphics[width=2.8in]{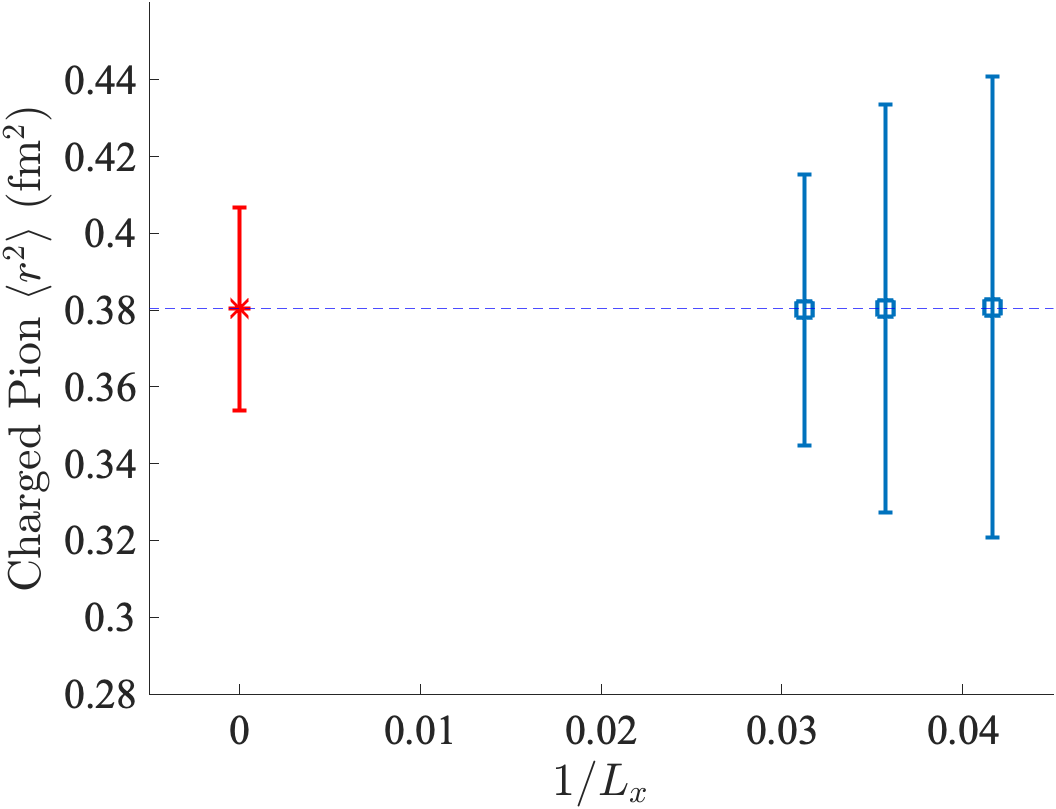} 
   \caption{Charge radius squared for $m_\pi = 315$ MeV (left) and $m_\pi = 220$ MeV (right). The blue points correspond to the different ensembles (see Table \ref{table:ensDesc}), and the red points are an average of the blue points for each mass. We have taken an average of the points rather than extrapolating to $1/L_x \rightarrow 0$ due to the size of the error bars. We expect only a small effect from the infinite-volume extrapolation.}
   \label{fig:r2}
\end{figure}

This procedure gives three data points as a function of volume for the $m_\pi = 315$ MeV case (Ensembles 1--3), and three data points for the $m_\pi = 220$ MeV case (Ensembles 4--6). The lattice elongation is in the x-direction, so in Fig. \ref{fig:r2} we have plotted $\langle r^2 \rangle$ as a function of inverse lattice size, $1/L_x$. At this point it is desirable to extrapolate to infinite volume, i.e., $1/L_x \rightarrow 0$. However, the error bars are too large to do a meaningful extrapolation. This is a result of the relatively large momenta used (we are working on analyzing ensembles at smaller momenta). Since the effect from the lattice size is expected to be small, the true result is approximated here as an average of the three points. It is worth emphasizing that these are preliminary results and we hope to reduce error bars by performing correlated fits on $Q_{44}^\text{elas}$ and analyzing smaller momenta.

\begin{figure}[h] 
   \centering
   \includegraphics[width=2.8in]{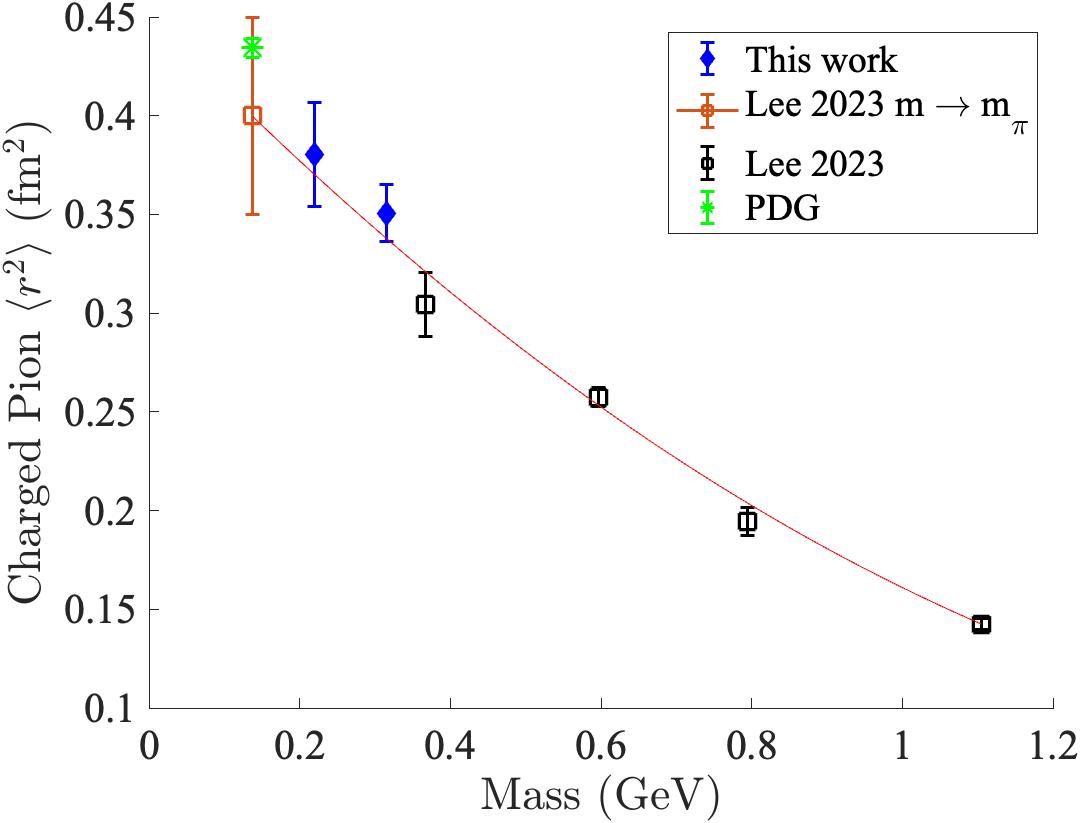} 
   \includegraphics[width=2.8in]{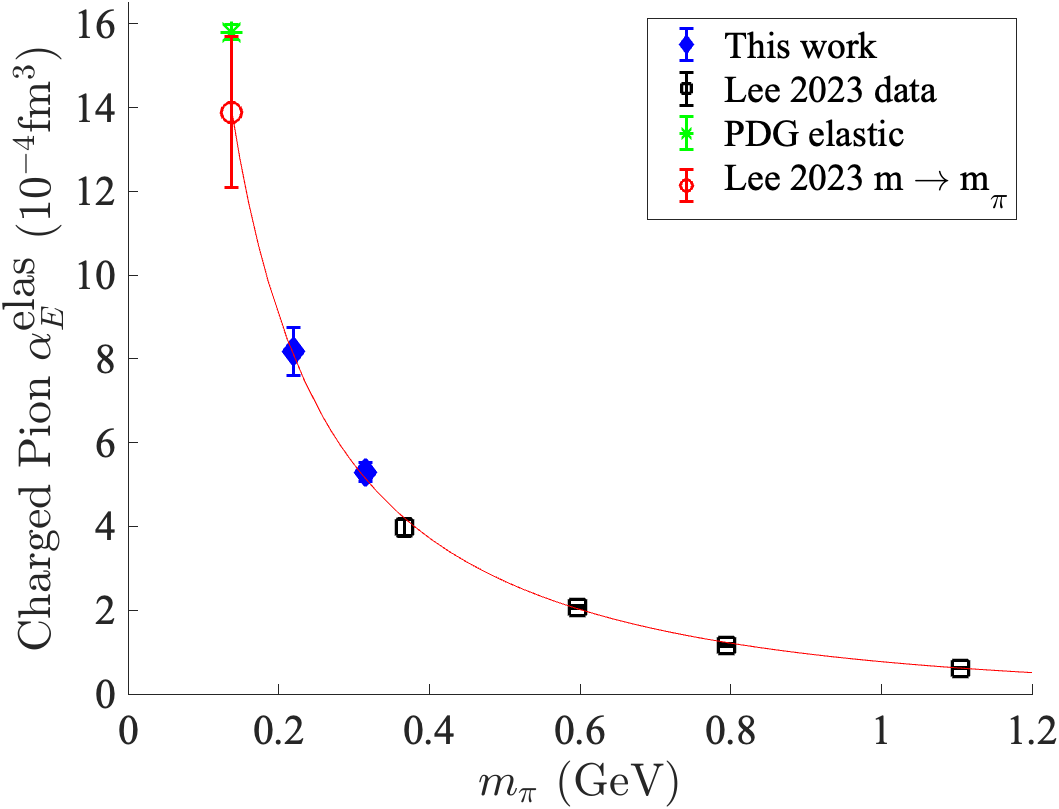} 
   \caption{Charge radius squared (left) and elastic contribution to the polarizability (right) from this study overlayed with with the same results from the results in Ref. \cite{Lee014512}. The solid blue diamond points correspond to the results from this work for $m_\pi = 315$ and $220$ MeV. Empty black square points are the results from Ref. \cite{Lee014512}. The red line is the fit of the four black points, which Lee et al. used to extrapolate to the physical pion mass. The green point is the result from the PDG.}
   \label{fig:r2Overlay}
\end{figure}

Having obtained data points for $\langle r^2 \rangle$ for masses $m_\pi= 220$ and $315$ MeV, we then compare these results to those obtained in the 2023 study \cite{Lee014512}. This is shown in Fig. \ref{fig:r2Overlay}, along with a comparison with the experimental value reported by the PDG. Though there is not a great deal of expectation that the results agree exactly (we use a different action), it is reassuring to see a high degree of compatibility.

The elastic contribution to the polarizability is now simple to calculate according to Eq.(\ref{eq:pol_elas}). The results are also shown in Fig. \ref{fig:r2Overlay}, overlayed with the results from Ref. \cite{Lee014512}. Once again, the agreement is very good, though we continue to moderate our expectations for agreement between analyses with different actions.

\subsection{Inelastic contribution}
\begin{figure}[h] 
   \centering
   \includegraphics[width = 3.0in]{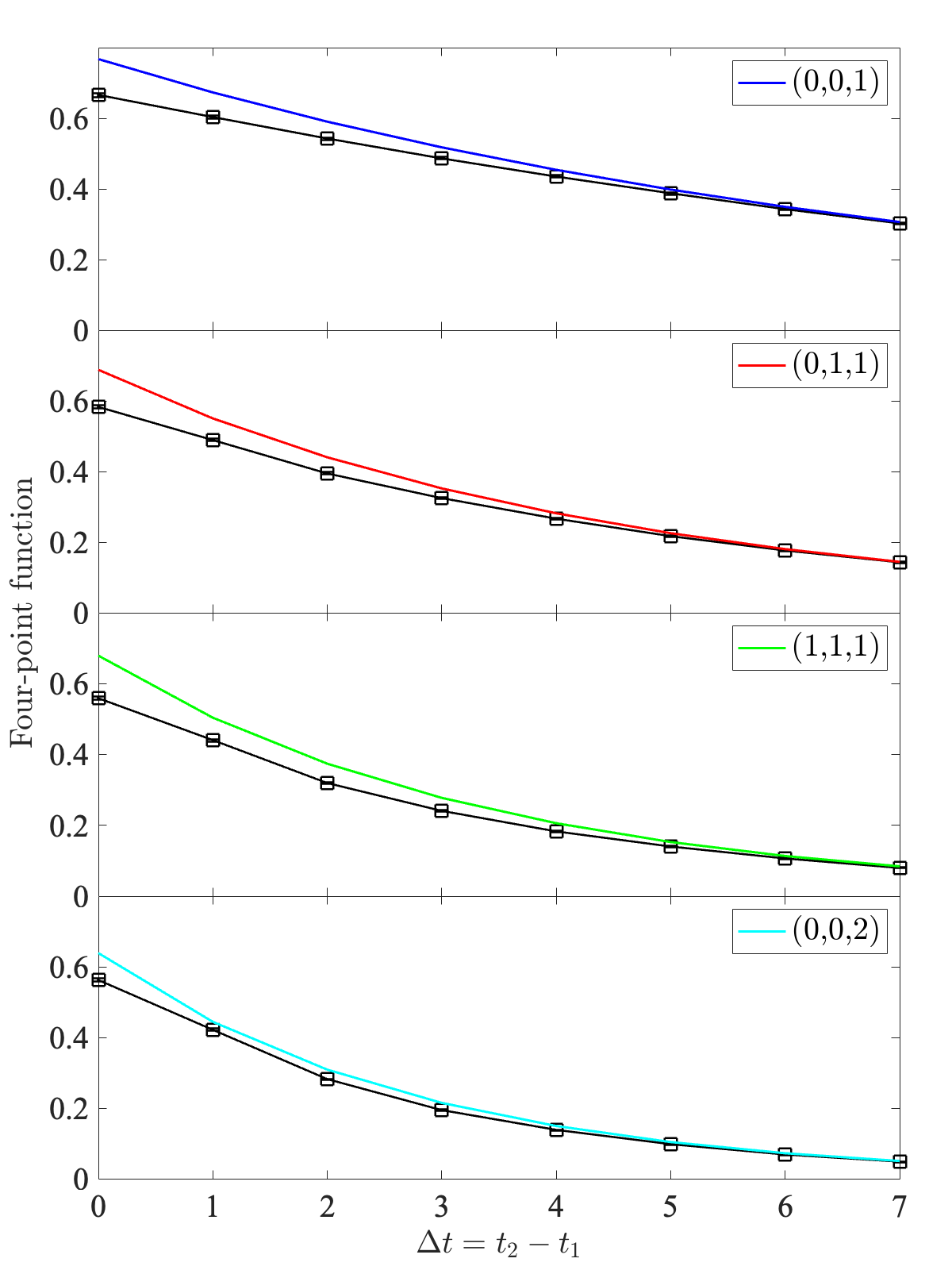} 
   \caption{Total four point function (sum of diagrams (a), (b), and (c), black markers), alongside the elastic fit $Q_{44}^\text{elas}$ (colored curves). We have only plotted the region of nonzero contribution. The area between the curves corresponds to the inelastic contribution.}
   \label{fig:4ptTotal}
\end{figure}

Now that the elastic contribution to the polarizability is computed, we turn our attention to the inelastic contribution, given by the integral expression in Eq.(\ref{eq:pol_inelas}). In words, Eq.(\ref{eq:pol_inelas}) tells us that the inelastic contribution to the polarizability is the area between the curves of the total four-point function (the sum of diagrams (a), (b), and (c)) and the elastic contribution to the four-point function. The good news is that $Q_{44}^\text{elas}$ was obtained in the previous section and will be used in the analysis of $\alpha_E^\text{inelas}$. Fig. \ref{fig:4ptTotal} shows clearly the area between the elastic and total four-point functions. It also suggests that the inelastic contribution will be negative. Note that Fig. \ref{fig:4ptTotal} is merely one representative sample of the four-point functions; the others look similarly.

We numerically compute the area between the curves. Although the point where $t_1 = t_2$ is unphysical, the time slice $0 < t_2 - t_1 < 1$ is the largest contribution to $\alpha_E^\text{inelas}$, so we include it in the analysis. To do this, we linearly extrapolate to $t_2 - t_1 = 0$ using the slope formed by the points $Q_{44}^{\text{ABC}}(t_2 - t_1 = 1,2)$. With the area found, the inelastic contribution is given by multiplying the area times $2\alpha/\qvec^2$. 

For each ensemble, we have four nonzero momenta, so that gives $\alpha_E^\text{inelas}$ as a function of momentum. Polarizability is a static property, so we smoothly extrapolate the data to $\qvec^2 \rightarrow 0$. We have elected to use two kinds of extrapolations to determine the reported value of $\alpha_E^\text{inelas}(\qvec^2 \rightarrow 0)$. The first is a quadratic form that uses all four data points. The second is a linear extrapolation using only the two smallest momenta. These values are averaged to obtain the reported value $\alpha_E^\text{inelas}$. The six ensembles yield six data points, shown on Fig. \ref{fig:polarEns}.

\subsection{Polarizability results}

\begin{figure}[h] 
   \centering
   \includegraphics[width=2.8in]{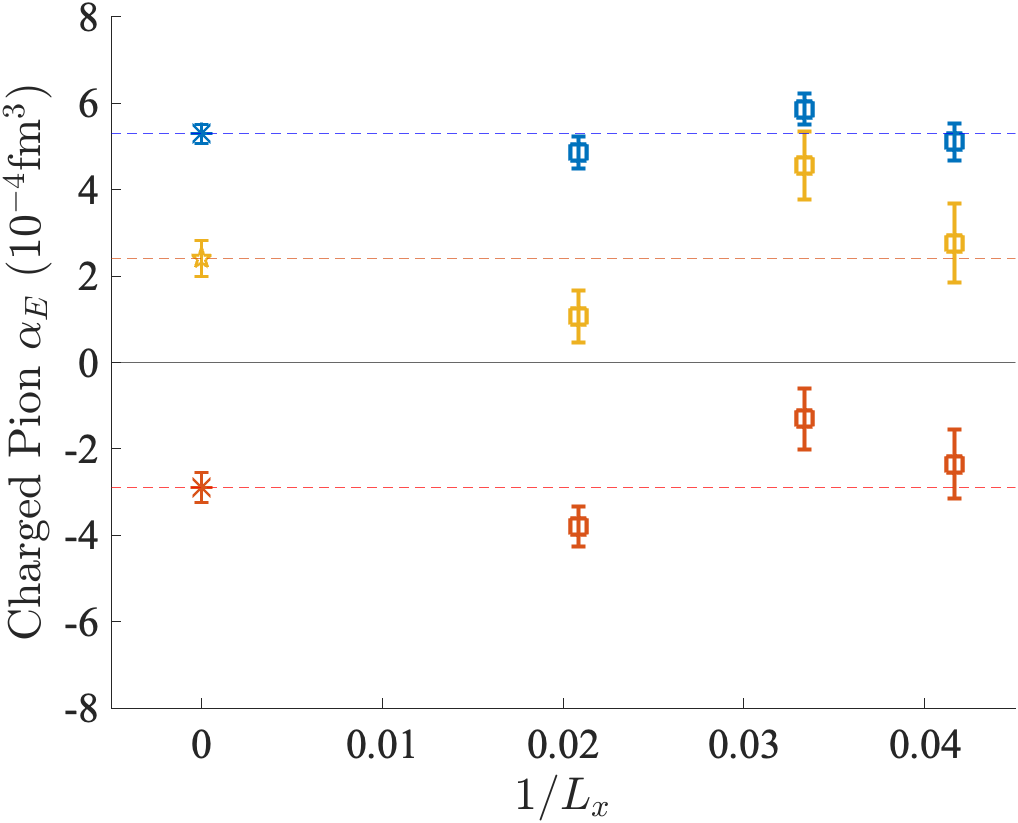} 
   \includegraphics[width=2.8in]{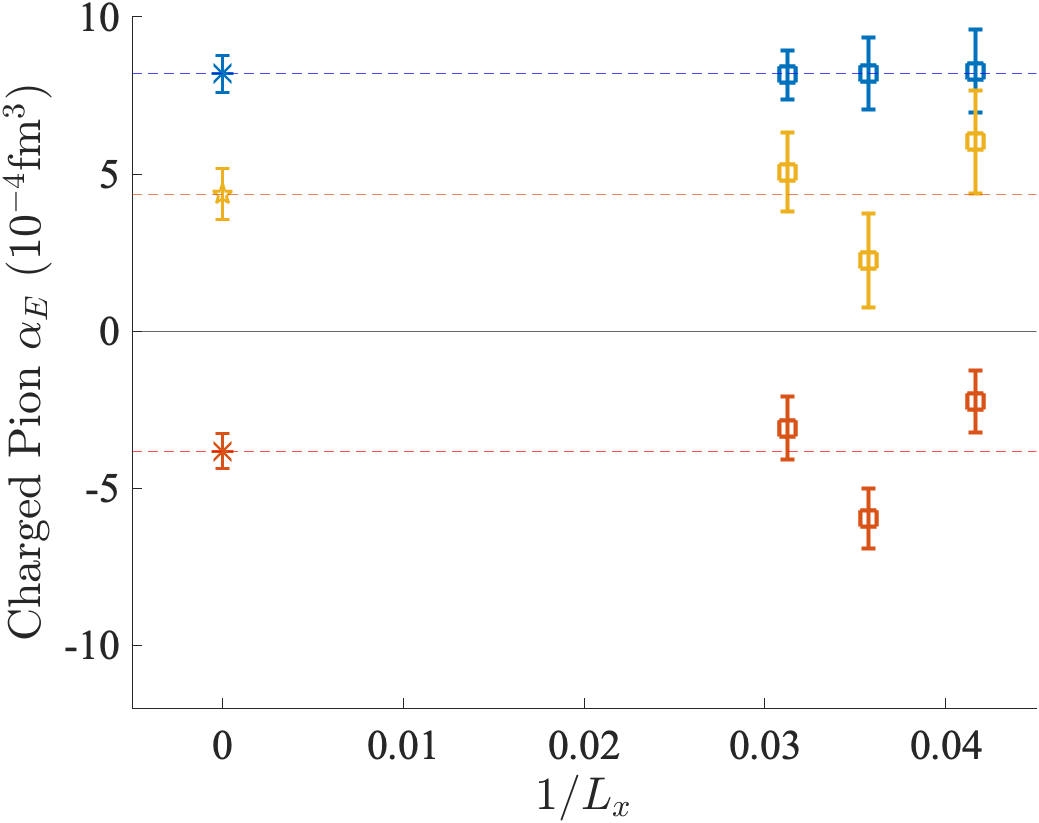} 
   \caption{Polarizability results corresponding to Ensembles 1 to 3 ($m_\pi = 315$ MeV, left) and Ensembles 4 to 6 ($m_\pi = 220$ MeV, right). Blue and red square markers represent elastic and inelastic contributions, respectively, and the gold square markers are the sum of the inelastic and elastic contributions. The blue and red asterisks are the averages of the elastic and inelastic contributions, respectively, and the gold star is the average of the total polarizability data points.}
   \label{fig:polarEns}
\end{figure}

Fig. \ref{fig:polarEns} shows the elastic and inelastic contributions to the polarizability, alongside the total, for each ensemble, plotted against inverse lattice size. Once again, the error bars are too significant to generate a meaningful extrapolation to infinite volume, so we approximate with an average, expecting small effects from an infinite-volume extrapolation. As was observed in Ref. \cite{Lee014512}, the elastic part $\alpha_E^\text{elas}$ contributes positively while the inelastic part $\alpha_E^\text{inelas}$ contributes negatively. The inelastic part is smaller in magnitude compared to the elastic part, yielding a positive total polarizability $\alpha_E^\pi$.

\begin{figure}[h] 
   \centering
   \includegraphics[width=3.4in]{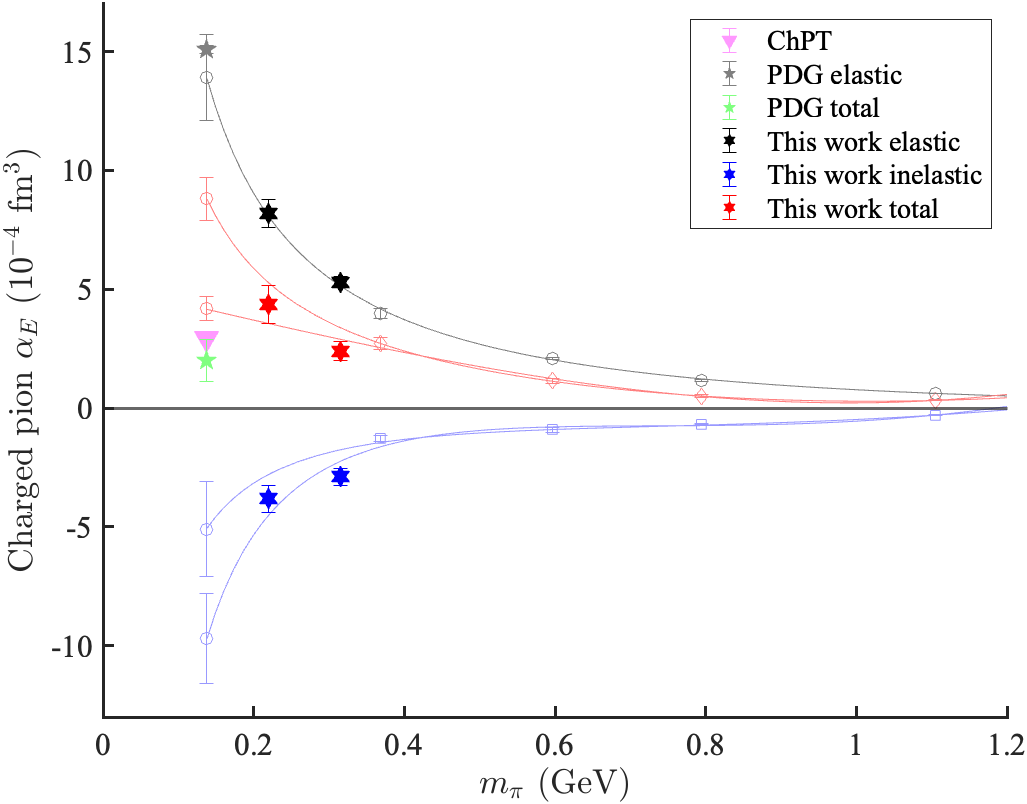} 
   \caption{Polarizability results from this study overlayed with the results from Ref. \cite{Lee014512} as a function of pion mass. The solid points in bold correspond to the results from this study, while the hollow points and extrapolated points at physical pion mass correspond to the results from Ref. \cite{Lee014512}. The results from chiral perturbation theory and the PDG are also included. Note there are two fit equations for the inelastic and total polarizability from the work of Lee et al. corresponding to upper and lower bounds.}
   \label{fig:polarAllOverlay}
\end{figure}

\begin{table}[h]\centering
\begin{tabular}{cccc} \hline\hline
$m_\pi$ (MeV) & Pion elastic $\alpha_E$ & Pion inelastic $\alpha_E$ & Pion $\alpha_E$ \\ \hline
315           & $5.30 \pm 0.22$           & $-2.9 \pm 0.4$            & $2.4 \pm 0.4$   \\
220           & $8.2 \pm 0.6$           & $-3.8 \pm 0.6$            & $4.4 \pm 0.8$  \\ \hline\hline
\end{tabular}
\caption{Elastic and inelastic contributions to the polarizability and the total polarizability results. These data points are plotted in Fig. \ref{fig:polarAllOverlay}. Polarizabilities are given in units of $10^{-4}$ fm$^3$.}
\end{table}

Finally, we compare these results with those reported in Ref. \cite{Lee014512} in Fig. \ref{fig:polarAllOverlay}. The data points marked in bold-colored stars are the averaged values from this study. As mentioned above, one of the advantages of this study over that of Ref. \cite{Lee014512} is that we use two smaller pion masses. It is encouraging to see that the new data points trend well with the existing results, filling in the ``mass gap'' at smaller masses. 

\section{Conclusion}
This study builds upon previous work by employing an alternate method to compute electric polarizabilities of charged pions using four-point functions. In particular, this study is advantageous over previous work because of the use of nHYP configurations, which offer dynamical fermions, smaller masses, and a volume dependence. 

The preliminary results we report show expected behavior when compared to the 2023 results of Lee et al \cite{Lee014512}. We highlight the fact that the total polarizability is understood as a partial cancellation between the inelastic and elastic terms, as has been previously observed. This cancellation appears to be true as we approach the physical point, though we have forgone an extrapolation to the physical mass since we examine only two masses. 

Another limitation at this stage in the analysis is that we have forgone an infinite-volume extrapolation due to the large error bars in the volume-dependent polarizability results. We would like to take advantage of the volume dependence of the ensembles in the future, and are in the process of extending this work to smaller momenta in order to reduce the error bars. Furthermore, the data is correlated, while at present we use uncorrelated fits. We anticipate that a correlated fit will further reduce the error bars. We also note that the results shown are only from the connected diagrams. A complete analysis would incorporate the disconnected diagrams; this is in the works also. 

Lastly, it is worth noting that the four-point function method we employ here can also be used for calculating neutral hadron polarizabilities \cite{Lee116701} and magnetic polarizabilities \cite{Lee054510}. We leave the calculation of these quantities for future work.

\section*{Acknowledgements}

This work was supported in part by U.S. Department of Energy under Grant No. DE-FG02-95ER40907. We would also like to acknowledge the Texas Advanced Computing Center (TACC) at The University of Texas at Austin as well as TAMU ACES and NCSA Delta resources through the NSF ACCESS program.

\bibliography{Lattice2025-refs.bib}

\end{document}